\documentclass[12pt, a4paper, twoside]{article}
 \usepackage[font=small,format=plain,labelfont=bf,up,textfont=normal,up,justification=justified,singlelinecheck=false]{caption}

\usepackage[a4paper, left=3cm,right=2cm]{geometry}
\usepackage{hyperref}
\usepackage{amsfonts}
\usepackage{amsmath}
\usepackage{setspace}
\usepackage{color}
\usepackage{multirow}

\usepackage[utf8,applemac]{inputenc}
\usepackage{tensor}
\usepackage{cite}
\usepackage{tikz}
\usepackage{graphicx}
\graphicspath{{figure/}}
\usepackage{dcolumn}
\usepackage{bm}
\usepackage[justification=centering]{caption} 

\newcommand{\bea}{\begin{eqnarray}}
\newcommand{\eea}{\end{eqnarray}}
\newcommand{\be}{\begin{equation}}
\newcommand{\ee}{\end{equation}}
\def\nn{\nonumber}

\def\p{\partial}

\newcommand{\cC}{\mathcal{C}}
\newcommand{\cD}{\mathcal{D}}
\newcommand{\cO}{\mathcal{O}}
\newcommand{\cT}{\mathcal{T}}

\newcommand{\Tr}{\textrm{Tr}}
\newcommand{\tr}{\textrm{tr}}

  \newcommand{\n}{\nabla }
  \newcommand{\beqs}{\begin{eqnarray}}
\newcommand{\eeqs}{\end{eqnarray}}

\def \a {\alpha}

\def \G {\Gamma}

\def \D {\Delta}
\def \e {\epsilon}

\def \m {\mu}
\def \n {\nu}

\def \l {\lambda}

\def \s {\sigma}

\def \r {\rho}

\def \th {\theta}

\def \p {\partial}
\def \f {\frac}

\def \nn {\nonumber}
\def \hs {\hspace}
\def \la {\langle}
\def \ra {\rangle}

\begin{document}
\title{R\'enyi Mutual Information for Free Scalar in Even Dimensions}
\author{
Bin Chen$^{1,2,3}$\footnote{bchen01@pku.edu.cn}\,
and
Jiang Long$^{4}$\footnote{Jiang.Long@ulb.ac.be}
}
\date{}

\maketitle

\begin{center}
{\it
$^{1}$Department of Physics and State Key Laboratory of Nuclear Physics and Technology,\\Peking University, 5 Yiheyuan Rd, Beijing 100871, P.R.\,China\\
\vspace{2mm}
$^{2}$Collaborative Innovation Center of Quantum Matter, 5 Yiheyuan Rd, \\Beijing 100871, P.R.\,China\\ \vspace{1mm}
$^{3}$Center for High Energy Physics, Peking University, 5 Yiheyuan Rd, \\Beijing 100871, P.R.\,China\\ \vspace{1mm}
$^{4}$Universit\'{e} Libre de Bruxelles and International Solvay Institutes,\\
CP 231, B-1050 Brussels, Belgium}
\vspace{10mm}
\end{center}

\begin{abstract}

 We compute the R\'enyi mutual information of two disjoint spheres in free massless scalar theory in even dimensions higher than two.  
 The spherical twist operator in a conformal field theory can be expanded into the sum of local primary operators and their descendants. 
 We analyze the primary operators in the replicated scalar theory and find the ones of the fewest dimensions and spins. 
 We study the one-point function of these operators in the conical geometry and obtain their expansion coefficients in the OPE of spherical twist operators. 
 We show that the R\'enyi mutual information can be expressed in terms of the conformal partial waves. 
 We compute explicitly the R\'enyi mutual information up to order $z^d$, where $z$ is the cross ratio and $d$ is the spacetime dimension.

\end{abstract}
\baselineskip 18pt
\thispagestyle{empty}

\newpage

\section{Introduction}

Entanglement entropy is an important notion in many body quantum system. It characterizes the entanglement between a subsystem with its environment.  It is defined as the von Neumann entropy of the reduced density matrix $\r_A$ of the subsystem $A$
\be
S_A=-\Tr_A \r_A\log \r_A.
\ee
 More generally, with the reduced density matrix, one may define the R\'enyi entanglement entropy  as
\be
S_A^{(n)}=-\f{1}{n-1} \log \Tr_A \r_A^n.
\ee
The R\'enyi entropy encodes more detailed spectral information on the reduced matrix. 
It is easy to see that the entanglement entropy can be obtained by taking the $n\to 1$ limit of the R\'enyi entropy 
\be
S_A=\lim_{n \to 1} S_A^{(n)}, \label{limit}
\ee
provided that such a limit is well-defined. 
One can also define the R\'enyi mutual information of two subsystems $A$ and $B$ as
\be
I_{A,B}^{(n)}=S_{A}^{(n)}+S_{B}^{(n)}-S_{A\cup B}^{(n)},
\ee
whose $n \to 1$ limit may define the mutual information
\be
I_{A,B}=S_{A}+S_{B}-S_{A\cup B}.
\ee
For a quantum system with finite number of degrees of freedom, the entanglement entropy can be computed numerically. 

The entanglement entropy in quantum field theory has been under active study since the pioneering work by Srednicki\cite{Srednicki:1993im}. As there are infinite number of degrees of freedom in a quantum field theory, the entanglement entropy is rather difficult to compute, even  numerically. It was surprisingly shown in \cite{Srednicki:1993im} that for free massless scalar, the entanglement entropy behaves like a geometric entropy as its leading contribution is proportional to the area of the boundary of the subsystem. Such area law of the entanglement entropy has shown to be quite generic in $d\geq 3$ spacetime\cite{Eisert:2008ur}. Intuitively it can be understood as originating from the correlations between the degrees of freedom on the different sides of the entangling surface.  In $d=2$, the entanglement entropy obeys a logarithmic law.  More importantly the single-interval entanglement and R\'enyi entropies in two-dimensional conformal field theory (CFT) display a universal behavior, being proportional to the central charge of the CFT\cite{Holzhey:1994we,Calabrese:2004eu}. The entanglement entropy in CFT has drown much interests in the past decade due to its relation with holography and quantum gravity\cite{Ryu:2006bv,Ryu:2006ef}. 

Despite much development in the past two decades, the computation of the entanglement entropy and its related quantities in quantum field theory is still a very hard problem\cite{Calabrese:2004eu,Calabrese:2005zw}. Most of the study has been focused on the simplest entangling region, the half space or a single sphere,  in the free theory\cite{Casini:2009sr} and 2D conformal field theory. 
From the relation (\ref{limit}), the entanglement entropy can be read from the R\'enyi entropy, which could be computed via the replica trick. However, this trick leads to the computation of the partition function on a spacetime manifold with singularity. For example, for a spherical entangling region, the replicated geometry is of conical singularity. For the case with multiple regions, the problem is even more difficult. 

The entanglement entropy of two disjoint regions $A$ and $B$ is interesting.  First of all, the mutual information between $A$ and $B$ measures the correlations between two regions. It is  positive, finite, free of ultra-violet divergence. In particular the mutual information satisfies the  inequality\cite{Wolf}
\be
I_{A,B}\geq \frac{\cC(M_A,M_B)^2}{2\parallel M_A\parallel^2 \parallel M_B\parallel^2},
\ee
where $M_A$ and $M_B$ are the observables in the regions $A$ and $B$ respectively, and $\cC(M_A,M_B):=\la M_A\otimes M_B \ra-\la M_A \ra \la M_B\ra $ is the connected correlation functions of $M_A$ and $M_B$. Holographically the classical mutual information of two well separated regions is vanishing, but the above inequality suggests that the quantum correction should give nonvanishing contribution\cite{Faulkner:2013ana}. In AdS$_3$/CFT$_2$, if one considers two intervals far apart  in the vacuum state of the large $c$ CFT, the leading contribution in the mutual information is vanishing but the subleading term independent of $c$ is not vanishing, which is in perfect match with 1-loop partition function of the graviton\cite{Barrella:2013wja, Chen:2013kpa}. 

It is a forbidding problem to compute the entanglement entropy or the mutual information of two disjoin regions directly via the  replica trick. The replicated geometry is not only of singularity but also of nontrivial topology. For example, in two-dimensional quantum field theory on complex plane, the $n$-th R\'enyi entropy of two intervals requires a partition function on a Riemann surface of genus $(n-1)$, which is hard to compute\cite{Calabrese:2009ez}. Nevertheless, when two intervals are short and far apart, one can use the operator product expansion(OPE) of the twist operators to compute the partition function order by order in the cross ratio\cite{Headrick:2010zt,Calabrese:2010he,Rajabpour:2011pt,Chen:2013kpa}. The OPE of the twist operators is of the form
\be
\s(z,\bar z)\tilde \s(0,0)
=c_n \sum_K d_K \sum_{m,r\geq0} \f{a_K^m}{m!}\f{\bar a_K^r}{r!}\f{1}{z^{2h-h_K-m}\bar z^{2\bar h-\bar h_K-r}} \p^m \bar \p^r \Phi_K(0,0),\nn
\ee
with the summation $K$ being over all the independent quasiprimary operators of  the replicated theory. The R\'enyi entropy of two intervals is the four-point function of twist operators in the orbifold CFT
\bea
 \Tr \r^n_A&=&\la \s(1+y,1+y) \tilde \s(1,1) \s(y,y)\tilde \s(0,0)  \ra_C  \nn\\
 &=&c_n^2 x^{-\f{c}{6}\left( n-\f{1}{n} \right)} \left( \sum_K \a_K d_K^2 x^{h_K} F(h_K,h_K;2h_K;x) \right)^2,\nn
\eea
where $x$ is the cross ratio and $F(h_K,h_K;2h_K;x)$ is the hypergeometric function. 
In principle, one can work out the partition function to any order. Especially for the large $c$ CFT dual to the AdS$_3$ gravity, the vacuum module contribution dominates the partition  function in the large central charge limit. This leads to a lot of study on the two-interval R\'enyi entropy in the large $c$ CFT, which sheds new light on the AdS$_3$/CFT$_2$ correspondence\cite{OPE}. 

The short interval expansion of the R\'enyi entropy can be applied to the CFT's in higher dimension. In \cite{Cardy:2013nua}, the mutual information of two disjoint spheres in the 3D and 4D massless scalar theory has been studied to the leading order of short interval expansion.  The result is consistent with the lattice computation in \cite{Shiba:2012np}  and the study in \cite{Casini:2008wt}. In \cite{Agon:2015twa}, the computation in 3D case has been pushed to the next leading order. For other study on the mutual information, see \cite{Agon:2015ftl,Schnitzer:2014zva,Herzog:2014fra}.

In this paper, we compute the mutual information of two disjoint spheres in free massless scalar field theory to higher orders and in other even dimensions. The key ingredient in our study is the OPE of spherical twist operator\cite{Hung:2014npa,Long:2016vkg,Bianchi:2015liz} in terms of the primary operators in the replicated theory. This allows us to write the R\'enyi mutual information in terms of conformal partial waves. We propose a partition function to count the number of independent non-descendant  operators,  work out explicitly the primary operators  of the first few lowest dimensions  and compute their one-point functions in the conical geometry to read the OPE coefficients. We 
manage to obtain the R\'enyi mutual information up to order $z^d$, where $z$ is the cross ratio and $d$ is the spacetime dimension.

The remaining parts of this paper are organized as follows. In the next section, we gave a brief review of spherical twist operator and its OPE expansion. In Section 3, we work out the primary operators in the replicated theory. In Section 4, we present the R\'enyi mutual information of two disjoint spheres in dimension 4, 6, 8 and 10. We end with conclusions and discussions in Section 5. Some technical details are put in a few appendices.

\section{Spherical twist operator}

The twist operators are naturally defined  through the replica trick. A twist operator is a co-dimension two operator and introduce the branch cut at the entangling surface in the path integral over the $n$-fold replicated theory. 
Let us focus on the case that the entangling surface $A$ is a single sphere of a radius $R$ and discuss the spherical twist operators.  In order to compute $\Tr \r_A^n$, one can compute the partition function of the CFT in the $n$-fold replicated geometry. Equivalently  one may introduce a spherical twist operator $\cT_n$ and compute the correlation function of the twist operator in the $n$-fold replicated theory. 
In two dimension, the twist operators are the local primary operator defined at the branch points. In higher dimensions, the twist operators are non-local surface operators. 

The twist operator in a conformal field theory is a defect operator of co-dimension two. Just like nonlocal Wilson loop or surface operators\cite{Shifman:1980ui,Berenstein:1998ij,Gomis:2009xg,Chen:2007zzr}, the defect operator can be expanded in 
terms of the local operators. In a conformal field theory, the local operators can be classified into primary fields and its descendants. For a defect operator\cite{Billo:2016cpy,Gadde:2016fbj} of co-dimension $q$ and  preserving $SO(d-q+1,1)\times SO(q-1,1)$ symmetry, it can be expanded into the sum of local operators
\be
\cD =<\cD>\sum_{\D,J}c_{\D,J}(\cO_{\D,J}+\mbox{its descendants}).
\ee
where $<\cD>$ is the expectation value of the defect operator and gives the partition function of the defect CFT. The coefficients $c_{\D,J}$ are determined by the one-point function of the operator $\cO_{\D,J}$ in the presence of the defect operator. Given a primary field, its descendant partners are fixed by the symmetry. Therefore, to understand a  defect operator it is essential to know the spectrum of the CFT and their one-point functions.   

For a spherical twist operator, 
it can be expanded in a similar way
\be
\cT_n=<\cT_n>\sum_{\D,J}c_{\D,J}Q[\cO_{\D,J}],
\ee
where $Q[\cO_{\D,J}]$ denotes all the operators generated from the primary field $\cO_{\D,J}$. 
 However the theory is $n$-fold replicated so that not only the primary fields in the original CFT have to be considered, the primary fields in different replica have to be taken into account. Shortly speaking, one has to consider the primary fields in CFT$_n$. Moreover, the one-point function of the primary field in the presence of the spherical twist operator can be 
equivalently computed by the one-point functions of the primary fields in the conical geometry. In general, such one-point functions may not be easy to compute. But for a free CFT, for example the free scalar we are interested in, these one-point functions change into multi-point function of free scalar in the conical geometry. 

In this work, we consider a CFT in its vacuum state in even dimensions  and  we let the entangling regions  be  spheres at a constant time slice. We consider  two disjoint spheres  such that the operator product expansion(OPE) of the twist operators can be applied. 
Let the two spheres be 
\be
A_1=\{t=0, \vec{x}^2\leq R^2\}, \hs{3ex}A_2=\{t=0, (\vec{x}-\vec{x}_0)^2\leq R'^2\}.
\ee
For simplicity, we use conformal symmetry to set $R=R'$ and $\vec{x}_0=(1,0,\cdots,0)$. The only independent conformal invariant quantity is the cross ratio
\be
z=\bar{z}=4R^2,\hs{5ex}u=z\bar{z},\hs{3ex}v=(1-z)(1-\bar{z}).
\ee
In the disjoint case, we have $0<z<1$. More generally, if the radii of two spheres are different, and their distance is $r$, then the cross ratio is 
\be
z=\bar{z}=\frac{4RR'}{r^2-(R-R')^2}. 
\ee

We are interested in the R\'enyi mutual information of two spheres. The R\'enyi mutual information between $A_1$ and $A_2$ is defined to be 
\be
I^{(n)}_{A_1,A_2}= S^{(n)}_{A_1}+S^{(n)}_{A_2}-S^{(n)}_{A_1\cup A_2}.
\ee
The R\'enyi entropy of $A_1\cup A_2$ is 
\be
S^{(n)}_{A_1\cup A_2}=\frac{1}{1-n}\log \la \cT_n(A_1\cup A_2)\ra. 
\ee
When the two regions are far apart, we may evaluate the expectation value of the twist operator approximately by 
\be
\frac{ \la \cT_n(A_1\cup A_2)\ra}{ \la \cT_n(A_1)\ra   \la \cT_n(A_2)\ra} =\frac{ \la \cT_n(A_1) \cT_n(A_2)\ra}{ \la \cT_n(A_1)\ra   \la \cT_n(A_2)\ra}= \sum_{\D,J}c^2_{\D,J}\la Q[\cO_{\D,J}](A_1)Q[\cO_{\D,J}](A_2) \ra.
\ee
The building block is the two-point function of the primary module, which is related to the conformal partial wave\footnote{The holographic description of conformal family has recently been studied from various points of view\cite{Czech:2016xec, deBoer:2016pqk,Chen:2016dfb}.}
\be
\la Q[\cO_{\D,J}](A_1)Q[\cO_{\D,J}](A_2) \ra \propto G_{\D,J}(u,v).
\ee
For example, for a scalar operator in four dimension, 
\be
\la Q[\cO_{\D,J=0}](A_1)Q[\cO_{\D,J=0}](A_2) \ra =N_{\D} \frac{\pi^42^{4-4\D}\G[\frac{\D}{2}-1]^4}{\G[\frac{\D-1}{2}]^2\G[\frac{\D+1}{2}]^2}G_{\D,J=0}(u,v).
\ee

In the end, we have the R\'enyi mutual information in terms of the conformal partial waves
\be
I^{(n)}_{A_1,A_2}=-\frac{1}{1-n}\log (1+\sum_{\D,J}s_{\D,J}G_{\D,J})
\ee
where 
\be
s_{\D,J}\sim\sum_{\D,J}c^2_{\D,J}\ee
 is the summation over all the primary operators with conformal dimension $\D$ and spin $J$.  It is zero when $n\to 1$, 
 \be
 s_{\D,J}(n=1)=0, 
 \ee
then one can expand it near $n=1$ as 
\be
s_{\D,J}=s'_{\D,J}(1)(n-1)+\frac{1}{2}s''_{\D,J}(1)(n-1)^2+\cdots.
\ee
The mutual information is just
\be
I_{A_1,A_2}=\sum_{\D,J}b_{\D,J}G_{\D,J}(u,v) , 
\ee
with 
\be
b_{\D,J}=s'_{\D,J}(1). 
\ee
This is the conformal block expansion of the mutual information. The coefficients $b_{\D,J}$ encode the dynamical information of corresponding conformal field theory. 

For the massless scalar theory, the strategy to compute the R\'enyi mutual information of two spheres is as follows. First, we have to work out the spectrum of the replicated theory. Next we need to work out the one-point functions of the primary operators in the presence of the twist operator, or multi-point function of the fields in the conical geometry. Finally, we consider the contributions of all the primary fields. In practice, it is hard to work out all the spectrum of the theory. In the small cross ratio limit, 
\be
G_{\Delta,J}^{(d)}(z)=z^{\Delta}+\cdots,
\ee
we can work out the R\'enyi mutual information in orders of the conformal dimension. 

\section{Primary operators in free scalar theory}

For a CFT in $d\geq 3$ dimension, the primary operators are classified by the scaling dimension and the representation under $SO(d-1,1)$ group. 
The conformal transformation requires that the Jacobian of a coordinate transformation $x\to x'$ must be proportional to the rotation matrix of the Lorenzian group 
\be
 J=\frac{\p x'^\m}{\p x^\n}=b(x)M^\m_{~\n}(x).
\ee
Under the conformal transformation, the primary operator is defined to be transformed as 
\be
\cO(x)\to \cO(x')=b(x)^{-\D}R[M^\m_{~\n}(x)]\cO(x),
\ee
where $\D$ is the scaling dimension of the operator, and $R$ is the spin representation of the operator under the Lorenzian group. By definition the primary operator at the origin is annihilated by the generator of special conformal transformation
\be
[\hat{K}_\m, \cO(0)]=0.
\ee

 To count the number of the primary operators with a fixed conformal dimension, it is convenient to use the partition function. The partition function for the free massless scalar theory can be defined to be\footnote{The partition function defined here is slightly different from the one defined by J. Cary in \cite{Cardy1991} and confirmed in \cite{Cardy:2016lei,Dowker:2016twk}. Here we count the types of the operators rather than the independent components. We clarify the difference of the partition function defined here and the one in \cite{Cardy1991,Cardy:2016lei,Dowker:2016twk} in Appendix D, and check the consistency of the two definitions to  the first few lowest dimensions. } 
\be
Z=\tr q^{\hat D}=\sum_{k\ge0}p(k)q^k,
\ee
where $\hat D$ is the dilatation operator and $p(k)$ count the number of the operators with scaling dimension $\Delta=k$. The number of primary operators with $\D=k$ is just\footnote{In principle, $m(k)$ is the number of independent types of operators which are not descendants.  One should decompose these operators into the irreducible representation of $SO(4)$. That means the number $m'(k)$ of the primary operators may be slightly larger than $m(k)$ in general, $m'(k)\ge m(k)$.  We checked up to level $6$ that $m'(k)=m(k)$ is valid for dimension 4. So we will not distinguish $m'(k)$ and $m(k)$ in this work. }
\be
m(k)=
p(k)-p(k-1),\ee 
 because that if $\cO_{k-1}$ is an operator with dimension $\D=k-1$, the operator $\p_\m \cO_{k-1}$ is of dimension $k$ but it is a descendent operator. Namely we have to remove the descendent operators being the derivative of the operators with lower dimension. 
 
 For a free massless scalar field theory, the building blocks of the primary operators are the free fields $\phi$ and the partial derivatives. Let us illustrate the construction in the four-dimensional theory.  
 Assume an operator is composed of $\phi$'s and  $\partial$'s,  whose numbers are $l$ and $k$ respectively, then its scaling dimension is $\Delta=k+l$. The problem is to count the number $f(k,l)$ of  independent  operators for a fixed dimension. There are relations
\be
f(k,l)=\left\{\begin{array}{ll}
f(k,l-1)+f(k-l,l),& k\ge l,\\
f(k,k),&k<l.\end{array}\right.
\ee
If there is no partial derivative, $k=0$, then there is only one independent operator $\phi^l$ for a fixed $l$
\be
f(0,l)=1,\ l\ge0.
\ee
For $k>0$, if there is no $\phi$ operator, $l=0$, then 
\be
f(k,0)=0.
\ee
We can count the number of  independent operators level by level, 
\be
f(0,0)=1,\ f(1,0)=0,f(0,1)=1,\ f(2,0)=0,\ f(1,1)=1,\ f(0,2)=1,\cdots.
\ee
Then 
\be
p(\D)=\sum_{k+l=\D; k,l \in \mathcal{Z}}f(k,l).
\ee
We find
\be
Z=1+q+2q^2+3q^3+5q^4+7q^5+11q^6+15q^7+\cdots.
\ee
The full partition function is 
\be
Z=\prod_{k=1}^{\infty}\frac{1}{1-q^k}.
\ee
In general $d$ dimension, the number of independent operators with dimension $\D$ is 
\be
p(\D)=\sum_{k+\frac{d-2}{2}l=\D;k,l\in\mathcal{Z}}f(k,l).
\ee
The partition function of free massless scalar in a general $d$ dimension could be 
\be
Z=\left\{\begin{aligned}&&\prod_{k\ge\frac{d-2}{2}}\frac{1}{1-q^k},\ d=4,6,\cdots,\\
&&\prod_{k\ge\frac{d-1}{2}}\frac{1}{1-q^{k-\frac{1}{2}}},\ d=3,5,\cdots.\end{aligned}\right.
\ee
 
In four dimension, when $\Delta= 0$, there is only one primary operator $I$. When $\Delta=1$, there is only one primary operator $\phi$.  When $\Delta=2$, there is only one primary operator $\phi^2$. When $\Delta=3$, there is only one primary operator $\phi^3$. When $\Delta=4$, there are two primary operators $\phi^4, T_{\mu\nu}$ after modulo the equation of motion. 
When $\D= 5$, there are two primary operators 
\be
\phi^5,\ \phi T_{\mu\nu}.
\ee
When $\D=6$, there are four primary operators
\be
\phi^6,\ \phi^2T_{\mu\nu},\ \phi_{\mu\nu\rho\sigma},\ \phi_{\mu\nu\rho},
\ee
where $\phi_{\mu\nu\rho\sigma}$ is a symmetric traceless spin-4 operator which is constructed by two $\phi$'s and four $\partial$'s, and $\phi_{\mu\nu\rho}$ is a symmetric traceless spin-3 operator which is constructed by three $\phi$'s and three $\partial$'s, 
\be
\phi_{\mu\nu\rho}=P^{\alpha\beta\gamma}_{\mu\nu\rho}(\partial_{\alpha}\phi\partial_{\beta}\phi\partial_{\gamma}\phi-\frac{1}{4}\phi(\partial_{\alpha}\partial_{\beta}\phi\partial_{\gamma}\phi+perm)+\frac{1}{12}\phi^2\partial_{\alpha}\partial_{\beta}\partial_{\gamma}\phi).
\ee
The projector is of the form
\be
P^{\alpha\beta\gamma}_{\mu\nu\rho}=\frac{1}{6}(\delta_{\mu}^{\alpha}\delta_{\nu}^{\beta}\delta_{\rho}^{\gamma}+perm.)-\frac{1}{18}(\delta^{\alpha\beta}(\delta_{\mu\nu}\delta^{\gamma}_{\rho}+\delta_{\mu\rho}\delta^{\gamma}_{\nu}+\delta_{\nu\rho}\delta^{\gamma}_{\mu})+perm.).
\ee
It is used to project a spin 3 operator to be symmetric and traceless. The primary operators up to dimension six are listed in Table 1. 
\begin{table}
 \centering
\begin{tabular}{|c|c|c|}
\hline
$\Delta$&primary operators&number\\
\hline
0 &$I$ &$1$ \\
\hline
1 & $\phi$&$1$ \\
\hline 2&$\phi^2$&$1$\\
\hline 3&$\phi^3$&$1$\\
\hline 4&$\phi^4$, $T_{\mu\nu}$&$2$\\
\hline 5&$\phi^5$,$ \phi T_{\mu\nu}$&$2$\\
\hline 6&$\phi^6$,$\phi^2T_{\mu\nu}$,$\phi_{\mu\nu\rho\sigma}$,$\phi_{\mu\nu\rho}$&$4$\\
\hline
\end{tabular}
\caption{Primary fields}
\end{table} 

For the $n$-fold replicated free theory,  the fields on different replica should be treated independently. 
The partition function is now
\be
Z_n=\tr q^{\sum_i {\hat D}_i}=\sum_{k\geq 0}p_n(k)q^k.
\ee
Similarly the number of the primary operators with dimension $k$ is given by 
\be
m_n(k)=p_n(k)-p_n(k-1).
\ee
In four dimension, we find 
\bea
Z_n&=&(1+q+2q^2+3q^3+5q^4+\cdots)^n\nn\\
&=&1+n q+\frac{1}{2}n(n+3)q^2+\frac{1}{6}n(n+1)(n+8)q^3+\frac{1}{24} n (1 + n) (3 + n) (14 + n) q^4\nn\\&&+\frac{1}{120} n (3 + n) (6 + n) (8 + n (21 + n)) q^5\nn\\
&& +\frac{1}{720} n (1 + n) (10 + n) (144 + n (181 + n (34 + n))) q^6+\cdots.
\eea

Now, we can find the primary operators of different dimensions. 
\begin{enumerate}
\item Dimension $\D=0$. There is only one primary operator: $I$.
\item Dimension $\D=1$. There are $n$ primary operators: $\phi_j$. 
\item Dimension $\D=2$. There are $\frac{n(n+3)}{2}-n=\frac{1}{2}n(n+1)$ primary operators: $\phi^2_j,\phi_{j_1}\phi_{j_2}$. 
\item Dimension $\D=3$. There are $\frac{1}{6}(n^2+6n-1)$ primary operators: $\phi^3_j,\phi^2_{j_1}\phi_{j_2},\phi_{j_1}\phi_{j_2}\phi_{j_3},J_{\mu}^{(j_1j_2)}$. Now there is a new type of  spin-1 operators 
\be
J_\m^{(j_1j_2)}=\phi_{j_1}\p_\m\phi_{j_2}-(j_1\leftrightarrow j_2),\hs{3ex}j_1>j_2, 
\ee
which do not appear in the single-copied theory. 
\item Dimension $\D=4$. There are $\frac{1}{24}n(n+1)(n^2+13n+10)$ primary operators, including $\phi^4_j,\phi^3_{j_1}\phi_{j_2},\phi^2_{j_1}\phi^2_{j_2},\phi^2_{j_1}\phi_{j_2}\phi_{j_3},\phi_{j_1}\phi_{j_2}\phi_{j_3}\phi_{j_4},J^{(1)(j_1j_2)}_{\mu},J^{(2)(j_1j_2j_3)}_{\mu},T^{(j)}_{\mu\nu},T^{(j_1j_2)}_{\mu\nu}$.
Here 
\bea
J^{(1)(j_1j_2)}_{\mu}&=&\phi_{j_1}\phi_{j_1}\partial_{\mu}\phi_{j_2}-\phi_{j_2}\phi_{j_1}\partial_{\mu}\phi_{j_1}=\phi_{j_1}J_{\mu}^{j_1j_2},\nn\\
J^{(2)(j_1j_2j_3)}_{\mu}&=&J_{\mu}^{(j_1j_2)}\phi_{j_3},\nn
\eea
and the stress tensor at the $j$-th replica
\be
T_{\mu\nu}^{(j)}=P^{\alpha\beta}_{\mu\nu}(\partial_{\alpha}\phi_j\partial_{\beta}\phi_j-\frac{1}{2}\phi_j\partial_{\alpha}\partial_{\beta}\phi_j)
\ee
with the projector 
\be
P^{\alpha\beta}_{\mu\nu}=\frac{1}{2}(\delta_{\mu}^{\alpha}\delta_{\nu}^{\beta}+\delta_{\nu}^{\alpha}\delta_{\mu}^{\beta})-\frac{1}{4}\delta_{\mu\nu}\delta^{\alpha\beta}, 
\ee
and 
\be
T_{\mu\nu}^{(j_1j_2)}=\frac{1}{2}P^{\alpha\beta}_{\mu\nu}[(\partial_{\alpha}\phi_{j_1}\partial_{\beta}\phi_{j_2}-\frac{1}{2}\phi_{j_1}\partial_{\alpha}\partial_{\beta}\phi_{j_2})+(j_1\leftrightarrow j_2)].
\ee
\end{enumerate}
The primary operators of different dimensions and degeneracies 
are listed in Table 2. When $\D=5$, there is a new kind of primary operator $\phi^{(j_1j_2)}_{\m\n\r}$ of the form
\be
\phi_{\mu\nu\rho}^{(j_1j_2)}=P^{\alpha\beta\gamma}_{\mu\nu\rho}(\phi_{j_1}\partial_{\alpha}\partial_{\beta}\partial_{\gamma}\phi_{j_2}-3\partial_{(\alpha}\phi_{j_1}\partial_{\beta}\partial_{\gamma)}\phi_{j_2}+3\partial_{(\alpha}\partial_{\beta}\phi_{j_1}\partial_{\gamma)}\phi_{j_2}-\phi_{j_2}\partial_{\alpha}\partial_{\beta}\partial_{\gamma}\phi_{j_1}).
\ee
The primary fields of free massless scalar in higher dimension can be discussed in a similar way.

\begin{table}
 \centering
\begin{tabular}{|c|c|c|}
\hline
$\Delta$&primary operators&number\\
\hline 0&$I$&$1$\\
\hline 1&$\phi_j$&$n$\\
\hline \multirow{2}{*}{2}&$\phi^2_j$&$n$\\
\cline{2-3}
&$\phi_{j_1}\phi_{j_2}$&$\frac{n(n-1)}{2}$\\
\hline\multirow{3}{*}{3}&$\phi^3_j$&$n$\\
\cline{2-3}
&$\phi_{j_1}^2\phi_{j_2}$&$n(n-1)$\\
\cline{2-3}
&$\phi_{j_1}\phi_{j_2}\phi_{j_3}$&$\frac{n(n-1)(n-2)}{6}$\\
\cline{2-3}
&$J_{\mu}^{(j_1j_2)}$&$\frac{n(n-1)}{2}$\\
\hline\multirow{9}{*}{4}&$\phi_j^4$&$n$\\
\cline{2-3}
&$\phi_{j_1}^3\phi_{j_2}$&$n(n-1)$\\
\cline{2-3}
&$\phi_{j_1}^2\phi_{j_2}^2$&$\frac{n(n-1)}{2}$\\
\cline{2-3}
&$\phi_{j_1}^2\phi_{j_2}\phi_{j_3}$&$\frac{n(n-1)(n-2)}{2}$\\
\cline{2-3}
&$\phi_{j_1}\phi_{j_2}\phi_{j_3}\phi_{j_4}$&$\frac{n(n-1)(n-2)(n-3)}{24}$\\
\cline{2-3}
&$\phi_{j_1}J_{\mu}^{(j_1j_2)}$&$n(n-1)$\\
\cline{2-3}
&$\phi_{j_1}J_{\mu}^{(j_2j_3)}$&$\frac{n(n-1)(n-2)}{3}$\\
\cline{2-3}
&$T_{\mu\nu}^{(j)}$&$n$\\
\cline{2-3}
&$T_{\mu\nu}^{(j_1j_2)}$&$\frac{n(n-1)}{2}$\\
\hline\multirow{14}{*}{5}&$\phi^5_j$&$n$\\
\cline{2-3}
&$\phi^4_{j_1}\phi_{j_2}$&$n(n-1)$\\
\cline{2-3}
&$\phi^3_{j_1}\phi^2_{j_2}$&$n(n-1)$\\
\cline{2-3}
&$\phi^3_{j_1}\phi_{j_2}\phi_{j_3}$&$\frac{n(n-1)(n-2)}{2}$\\
\cline{2-3}
&$\phi^2_{j_1}\phi_{j_2}^2\phi_{j_3}$&$\frac{n(n-1)(n-2)}{2}$\\
\cline{2-3}
&$\phi^2_{j_1}\phi_{j_2}\phi_{j_3}\phi_{j_4}$&$\frac{n(n-1)(n-2)(n-3)}{6}$\\
\cline{2-3}
&$\phi_{j_1}\phi_{j_2}\phi_{j_3}\phi_{j_4}\phi_{j_5}$&$\frac{n(n-1)(n-2)(n-3)(n-4)}{120}$\\
\cline{2-3}
&$\phi_{j_1}\phi_{j_2}J_{\mu}^{(j_1j_2)}$&$\frac{n(n-1)}{2}$\\
\cline{2-3}
&$\phi_{j_1}^2J_{\mu}^{(j_1j_2)}$&$n(n-1)$\\
\cline{2-3}
&$\phi_{j_1}^2J_{\mu}^{(j_2j_3)}$ and $\phi_{j_1}\phi_{j_2}J_{\mu}^{(j_1j_3)}$&$n(n-1)(n-2)$\\
\cline{2-3}
&$\phi_{j_1}\phi_{j_2}J_{\mu}^{(j_3j_4)}$&$\frac{n(n-1)(n-2)(n-3)}{8}$\\
\cline{2-3}
&$\phi_jT_{\mu\nu}^{(j)}$&$n$\\
\cline{2-3}
&$\phi_{j_1}T_{\mu\nu}^{(j_2)}$&$n(n-1)$\\
\cline{2-3}
&$\phi_{j_1}T_{\mu\nu}^{(j_1j_2)}$&$n(n-1)$\\
\cline{2-3}
&$\phi_{j_1}T_{\mu\nu}^{(j_2j_3)}$&$\frac{n(n-1)(n-2)}{2}$\\
\cline{2-3}
&$\phi_{\mu\nu\rho}^{(j_1j_2)}$&$\frac{n(n-1)}{2}$\\
\hline
\end{tabular}
\caption{Primary fields in $n$-fold replicated theory}
\end{table}

\section{R\'enyi mutual information of free scalar}
In this section, we compute the R\'enyi mutual information between two spheres for the massless scalars in even dimensions. 
We have discussed the primary fields in the $n$-fold replicated theory and obtained the first few low-dimensions in the last section. The missing piece 
is the coefficients of the one-point function of the primary fields in the presence of the twist field or the one-point function in the conical geometry. It can be computed by the multi-point functions of scalar fields in the conical geometry. For a scalar primary field of scaling dimension $\D$, its one-point function in a conical geometry should be of the form 
\be
\la \cO(x) \ra =\frac{a_{\D,0}}{|x|^\D},
\ee
where $a_{\D,0}$ is the coefficients. For a general symmetric traceless primary field of scaling dimension $\D$ and spin $J$, its one-point function is fixed up to a constant $a_{\D,J}$. We collect the one-point function of spin-1 and spin-2 operators in various even dimensions in Appendix B. 
With the normalization $G^{(d)}_{\D,J}\sim z^{\D}$ in the small cross ratio limit, the coefficients $s_{\D,J}$'s are related to $a_{\D,J}$'s by 
\be
s_{\D,0}=\sum_{\mathcal{O}_{\D,0}}\frac{a^2_{\D,0}}{N_{\D,0}},\  s_{\D,1}=-\sum_{\mathcal{O}_{\D,1}}\frac{a^2_{\D,1}}{N_{\D,1}},\  s_{\D,2}=d(d-1)\sum_{\mathcal{O}_{\D,2}}\frac{a^2_{\D,2}}{N_{\D,2}}.
\ee
The summation is over all primary fields with dimension $\D$ and  spin $J$.  One can also derive the expression of the $s_{\D,J}$ for higher spin fields, but we will not need them in this work. 
For the free scalar theory, it is obvious that only the primary fields including even number of the scalars have nonvanishing one-point function. 

\subsection{Four dimensional case}
We first work on the $d=4$ free scalar theory. The two-point function of scalar in flat spacetime is 
\be
<\phi(x)\phi(x')>=\frac{1}{(x-x')^2}.
\ee
In the conical geometry it is of the form \cite{Guimaraes:1994,Nozaki:1401}
\be
<\phi(x_1)\phi(x_2)>_n=\frac{G_n(x_1,x_2)}{n},
\ee
where 
\be
G_n(x_1,x_2)=\frac{1}{2r_1r_2}\frac{\sinh\frac{\eta_{12}}{n}}{\sinh\eta_{12}(\cosh\frac{\eta_{12}}{n}-\cos\frac{\theta_{12}}{n})}.
\ee
Here we have used  Euclidean coordinates $x=(r,\theta,\vec{y})$ and have defined 
\be
\theta_{12}=\theta_1-\theta_2,\hs{3ex} \cosh\eta_{12}=\frac{r_1^2+r_2^2+ \vec{y}_{12}^2}{2r_1r_2}.
\ee
The scalar four-point function is 
\bea
\lefteqn{<\phi(x_1)\phi(x_2)\phi(x_3)\phi(x_4)>_n}\nn\\
&=&\frac{1}{n^2}(G_n(x_1,x_2)G_n(x_3,x_4)+G_n(x_1,x_3)G_n(x_2,x_4)+G_n(x_1,x_4)G_n(x_2,x_3)).
\eea

The first nonvanishing contribution to the R\'enyi mutual information comes from the dimension-2 operators: 
\be
\cO_{2,0}^{(1)}=\phi_{j}^2, \hs{3ex}\cO_{2,0}^{(2)}=\phi_{j_1}\phi_{j_2}, \hs{1ex}(j_1> j_2).
\ee
Their one-point coefficients and normalization factors are respectively 
\bea
a_{2,0}^{(1)}=\frac{1-n^2}{12n^2}, &&N_{2,0}^{(1)}=2,\nn\\
a_{2,0}^{(2)}=\frac{1}{4n^2\sin^2\frac{\th_{ij}}{2n}}, &&N_{2,0}^{(2)}=1.
\eea
Then we have 
\be
s_{2,0}=\frac{n}{2}\left(\frac{1-n^2}{12n^2}\right)^2+\sum_{j_1>j_2}\left(\frac{1}{4n^2\sin^2\frac{\th_{ij}}{2n}}\right)^2=\frac{n^4-1}{240n^3}. 
\ee

The next nonvanishing contribution comes from the dimension-3 operators of spin 1
\be
\cO_{3,1}=J_\m^{(j_1j_2)},
\ee
which has 
\be
a_{3,1}=\frac{\cos\frac{\th_{ij}}{2n}}{2n^3\sin^3\frac{\th_{ij}}{2n}}, \hs{3ex}N_{3,1}=4. 
\ee 
So we have\footnote{Note we use a different normalization of the conformal block compared to \cite{Long:2016vkg}.}
\be
s_{3,1}=\sum_{i>j}\frac{1}{4}\left(\frac{\cos\frac{\th_{ij}}{2n}}{2n^3\sin^3\frac{\th_{ij}}{2n}}\right)^2=\frac{n^6-21n^2+20}{15120n^5}.
\ee
Here we have taken into account of the fact that the Wick rotation on the spin-1 operators $J_\m^{(j_1j_2)}$  contribute a minus sign in the final coefficient\footnote{We would like to thank Lin Chen for pointing out the contribution from the Wick rotation.}. 

The  primary operators of dimension 4 giving nonvanishing contribution consist of two classes: the ones without spin and the ones with spin 2.  The ones without spin include the following types of the operators 
\be
\phi^4_j,\ \phi^3_{j_1}\phi_{j_2},\ \phi^2_{j_1}\phi^2_{j_2},\ \phi^2_{j_1}\phi_{j_2}\phi_{j_3},\ \phi_{j_1}\phi_{j_2}\phi_{j_3}\phi_{j_4}.
\ee
Their one-point function coefficients are respectively 
\bea
&&a_j=\frac{(1-n^2)^2}{48n^4},\ a^{(1)}_{j_1j_2}=\frac{1-n^2}{16n^4\sin^2\frac{\theta_{j_1j_2}}{2n}},\ a^{(2)}_{j_1j_2}=\frac{(1-n^2)^2}{144n^4}+\frac{1}{8n^4\sin^4\frac{\theta_{j_1j_2}}{2n}},\nn\\
&&a_{j_1j_2j_3}=\frac{1-n^2}{48n^4\sin^2\frac{\theta_{j_2j_3}}{2n}}+\frac{1}{8n^4\sin^2\frac{\theta_{j_1j_2}}{2n}\sin^2\frac{\theta_{j_1j_3}}{2n}},\\
&&a_{j_1j_2j_3j_4}=\frac{1}{16n^4}(\frac{1}{\sin^2\frac{\theta_{j_1j_2}}{2n}\sin^2\frac{\theta_{j_3j_4}}{2n}}+\frac{1}{\sin^2\frac{\theta_{j_1j_3}}{2n}\sin^2\frac{\theta_{j_2j_4}}{2n}}+\frac{1}{\sin^2\frac{\theta_{j_1j_4}}{2n}\sin^2\frac{\theta_{j_2j_3}}{2n}}).\nn
\eea
And their normalization factors are respectively
\be
N_j=24,\ N^{(1)}_{j_1j_2}=6,\ N^{(2)}_{j_1j_2}=4,\ N_{j_1j_2j_3}=2,\ N_{j_1j_2j_3j_4}=1.
\ee
Then we find 
\be
s_{4,0}=\frac{-420+7n+400n^2-14n^5+20n^8+7n^9}{806400n^7}.
\ee
The other class includes the spin-2 operators 
\be
T_{\mu\nu}^{(j)}, \hs{4ex}T_{\m\n}^{(j_1j_2)}\hs{1ex}(j_1> j_2).
\ee
Their one-point function coefficients and normalization factors are respectively 
\bea
a^{(j)}_T=\frac{1-n^4}{240n^4}, & &N_T^{(j)}=12, \nn\\
a^{(j_1j_2)}_T=\frac{2+\cos\frac{\theta_{j_1j_2}}{n}}{16n^4\sin^4\frac{\theta_{j_1j_2}}{2n}},&&N^{(j_1j_2)}_{T}=6.
\eea
Then we get 
\be
s_{4,2}=\frac{-21+20n^2+n^8}{40320n^7}.
\ee
It is possible to consider the contributions from the primary operators of higher dimension, but the computation becomes more involved. 

With all the information above, we can read the R\'enyi mutual information 
\bea
I^{(n)}_{A_1,A_2}&=&\frac{1}{n-1}\log\left(1+\frac{n^4-1}{240n^3}G_{2,0}(z)+\frac{n^6-21n^2+20}{15120n^5}G_{3,1}(z)\right.\\&&\left.+\frac{-420+7n+400n^2-14n^5+20n^8+7n^9}{806400n^7}G_{4,0}(z)+\frac{-21+20n^2+n^8}{40320n^7}G_{4,2}(z)+\cdots\right)\nn
\eea
In the small cross ratio limit, we have 
\bea
I^{(n)}_{A_1,A_2}&=&\frac{(1+n)(1+n^2)}{240n^3}z^2+\frac{(1 + n) (-1 + 2 n) (1 + 2 n) (5 + 4 n^2)}{3780 n^5}z^3+\nn\\&&\frac{(1 + n) (7 - 13 n^2 + 29 n^4 + 29 n^6)}{6720 n^7}z^4+\cdots\nn
\eea
The mutual information is 
\be
I_{A_1,A_2}=\frac{1}{60}G_{2,0}(z)-\frac{1}{420}G_{3,1}(z)+\frac{1}{840}G_{4,0}(z)+\frac{1}{840}G_{4,2}(z)+\cdots
\ee
In the small cross ratio limit, it is
\be
I_{A_1,A_2}=\frac{1}{60}z^2+\frac{1}{70}z^3+\frac{13}{840}z^4+\cdots.
\ee
The leading contribution in the small $z$ agrees with the result in \cite{Cardy:2013nua,Shiba:2012np}.


\subsection{Other dimensions}

In other dimensions, the discussion is similar. The twist operator of two spheres can be expanded by the OPE, which leads to 
\be
\frac{<\cT_{A\cup B}>}{<\cT_A><\cT_B>}\sim 1+\sum_{\Delta,J}\frac{a_{\Delta,J}^2}{N_{\Delta,J}}G_{\Delta,J}^{(d)}(z)+\cdots.
\ee
The  R\'enyi mutual information  is
\bea
I^{(n)}_{A_1,A_2}&=&\frac{1}{n-1}\log\left(1+\sum_{\Delta,J=0}\frac{a_{\Delta,J=0}^2}{N_{\Delta,J=0}}G_{\Delta,J=0}^{(d)}(z)+\sum_{\Delta,J=1}(-1)\frac{a_{\Delta,J=1}^2}{N_{\Delta,J=1}}G_{\Delta,J=1}^{(d)}(z)\right.\nn\\
&&\left. +\sum_{\Delta,J=2}d(d-1)\frac{a_{\Delta,J=2}^2}{N_{\Delta,J=2}}G_{\Delta,J=2}^{(d)}(z)+\cdots\right).
\eea
In odd dimensions, the scalar two-point function in the conical space has no analytic form. In even dimensions, such two-point functions are known. We collect them in Appendix A.  In this paper, we focus on the free scalar in  even dimensions.

The primary operators in the replicated theory can be discussed as in four dimensional case. 
In dimension $d$, a free scalar is of scaling dimension 
\be
[\phi]=\frac{d-2}{2}. 
\ee
Up to the scaling dimension $d$, the primary operators with nonvanishing contributions are similar to the ones in $d=4$. They include the operators without spin
\be
\phi_j^2, \hs{2ex}\phi_{j_1}\phi_{j_2},
\ee
the spin-1 operator 
\be
J_\m^{(j_1j_2)},
\ee
and the spin-2 operator
\be
T_{\m\n}^{(j)},\hs{3ex}T_{\m\n}^{(j_1j_2)}. 
\ee
Now the 
stress tensor in general dimension is 
\be
T_{\mu\nu}^j=P^{\alpha\beta}_{\mu\nu}(\partial_{\alpha}\phi\partial_{\beta}\phi-\frac{d-2}{d}\phi\partial_{\alpha}\partial_{\beta}\phi)
\ee
where the projector is 
\be
P^{\alpha\beta}_{\mu\nu}=\frac{1}{2}(\delta^{\alpha}_{\mu}\delta^{\beta}_{\nu}+\delta^{\alpha}_{\nu}\delta^{\beta}_{\mu})-\frac{1}{d}\delta_{\mu\nu}\delta^{\alpha\beta}.
\ee
The operator $T_{\m\n}^{(j_1j_2)}$ is 
\be
T_{\m\n}^{j_1j_2}=\frac{1}{2}P^{\alpha\beta}_{\mu\nu}(\partial_{\alpha}\phi^{j_1}\partial_{\beta}\phi^{j_2}-\frac{d-2}{d}\phi^{j_1}\partial_{\alpha}\partial_{\beta}\phi^{j_2}+(j_1\leftrightarrow j_2)).
\ee
The normalization factors of $T$'s are respectively
\be
N_T^j=\frac{4(d-1)(d-2)^2}{d},\hs{3ex} N_{T}^{j_1j_2}=\frac{2(d-1)(d-2)^2}{d}.
\ee

The coefficients of the one-point functions and the normalization factors are different in different dimensions. We collect them in Appendix B. 
Here we just give the  R\'enyi mutual information in different dimensions. In $d=6$, we have 
\bea
I^{(n),d=6}_{A_1,A_2}&=&\frac{1}{n-1}\log\left(1-\frac{21+40n^2+42n^4-103n^8}{725760n^7}G_{4,0}^{(d=6)}(z)\right.\nn\\&&+\frac{420+231n^2-220n^4-462n^6+31n^{10}}{19958400n^9}G_{5,1}^{(d=6)}(z)\nn\\&&\left. +\frac{-1382+819n^4 + 520 n^6 + 43 n^{12}}{113218560n^{11}}G_{6,2}^{(d=6)}(z)+\cdots\right).
\eea
When $z<<1$, the leading terms in the R\'enyi mutual information are
\bea
I^{(n),d=6}_{A_1,A_2}&=&\frac{(n+1)(21 + 61 n^2 + 103 n^4 + 103 n^6) }{725760 n^7}z^4\nn\\&&+\frac{(1 + n) (-1 + 2 n) (1 + 2 n) (105 + 294 n^2 + 445 n^4 + 356 n^6)}{4989600 n^9}z^5\nn\\&&+\frac{(1 + n)(15202 - 50318 n^2 + 12745 n^4 + 247265 n^6 + 535553 n^8 + 535553 n^{10})}{1245404160 n^{11}}z^6+\cdots\nn
\eea
The mutual information is 
\be
I^{d=6}_{A_1,A_2}=\frac{1}{1260}G_{4,0}^{(d=6)}(z)-\frac{1}{6930}G_{5,1}^{(d=6)}(z)+\frac{1}{16380}G_{6,2}^{(d=6)}(z)+\cdots.
\ee
In the small cross ratio limit, it is approximated by 
\be
I^{d=6}_{A_1,A_2}=\frac{1}{1260}z^4+\frac{1}{693}z^5 + \frac{25}{12012}z^6+\cdots.
\ee

In $d=8$, we have the R\'enyi mutual information
\be
I^{(n),d=8}_{A_1,A_2}=\frac{1}{n-1}\log(1+s^{(d=8)}_{6,0} G_{6,0}^{(d=8)}(z)+s^{(d=8)}_{7,1} G_{7,1}^{(d=8)}(z)+s^{(d=8)}_{8,2} G_{8,2}^{(d=8)}(z)+\cdots),
\ee
with 
\bea
s^{(d=8)}_{6,0}&=&-\frac{15202+54600n^2+99099 n^4+114400n^6+96096n^8-379397n^{12}}{62270208000n^{11}},\nn\\
s^{(d=8)}_{7,1}&=&\frac{120120 + 258434 n^2 + 196560 n^4 - 57057 n^6 - 251680 n^8 - 
 288288 n^{10} + 21911 n^{14}}{435891456000 n^{13}},\nn\\
s^{(d=8)}_{8,2}&=&\frac{-3620617 - 4764760 n^2 - 516868 n^4 + 3650920 n^6 + 3539536 n^8 + 
 1555840 n^{10} + 155949 n^{16}}{16937496576000 n^{15}}.\nn
\eea
When $z<<1$, the leading terms  are
\be
I^{(n),d=8}_{A_1,A_2}=\alpha_6 z^6+\alpha_7 z^7+\alpha_8 z^8+\cdots
\ee
where
\bea
\alpha_6&=&\frac{(1 + n) (15202 + 
69802n^2 + 168901  n^4+ 283301 n^6 + 379397 n^8 + 379397 n^{10})}{62270208000 n^{11}},\nn\\
\alpha_7&=&\frac{(1 + n) (-1 + 2 n) (1 + 2 n)}{54486432000 n^{13}}(15015 + 67474 n^2 + 158555 n^4 + 255612 n^6 + 312080 n^8 \nn\\&&+ 249664 n^{10})\nn\\
 \alpha_8&=&\frac{(1+n)}{16937496576000 n^{15}}(3620617 - 7950943 n^2 - 17771435 n^4 + 40952685 n^6 + 206902469 n^8 + \nn\\&&
 426275909 n^{10} + 622311749 n^{12} + 622311749 n^{14})
\eea
The mutual information is 
\be
I^{d=8}_{A_1,A_2}=\frac{1}{24024}G_{6,0}^{(d=8)}(z)-\frac{1}{120120}G_{7,1}^{(d=8)}(z)+\frac{41}{12252240}G_{8,2}^{(d=8)}(z)+\cdots
\ee
When $z<<1$, it is approximated by 
\be
I^{d=8}_{A_1,A_2}=\frac{1}{24024}z^6 + \frac{1 }{8580 }z^7+ \frac{49} {218790}{z^8}+\cdots.
\ee

In $d=10$, we  get  the R\'enyi mutual information 
\bea
I^{(n),d=10}_{A_1,A_2}&=&\frac{1}{n-1}\log(1+s_{8,0}^{(d=10)}G_{8,0}^{(d=10)}(z)+s_{9,1}^{(d=10)}G_{9,1}^{(d=10)}(z)+s_{10,2}^{(d=10)}G_{10,2}^{(d=10)}(z)+\cdots)\nn
\eea
where 
\bea
s_{8,0}^{(d=10)}&=&-\frac{1}{237124952064000n^{15}}(517231+2722720n^2+7236152n^4+12764960n^6\nn\\&&
+16574558n^8+16336320n^{10}+12602304n^{12}-68754245n^{16}),\nn\\
s_{9,1}^{(d=10)}&=&\frac{1}{638636777146368000 n^{17}}(2132813540 + 8048631591 n^2 + 13851157320 n^4 + \nn\\&&12609511728 n^6 + 
  2756475540 n^8 - 9573594030 n^{10} - 15962918400 n^{12}\nn\\&& - 
  15084957888 n^{14} + 1222880599 n^{18}),\nn\\
  s_{10,2}^{(d=10)}&=&\frac{1}{9461285587353600000 n^{19}}(-30794046738 - 85312541600 n^2 - 93900701895 n^4\nn\\&& - 25995169200 n^6 + 
  58112761410 n^8 + 89131333200 n^{10} \nn\\&&+ 62155613520 n^{12} + 
  23944377600 n^{14} + 2658373703 n^{20}).\nn
\eea
When $z<<1$, it is approximated by 
\bea
I^{(n)}_{A_1,A_2}&=&\beta_8z^8+\beta_9z^9+\beta_{10}z^{10}+\cdots.
\eea
with
\bea
\beta_8&=&\frac{ 1 + n}{237124952064000 n^{15}}(517231 + 3239951 n^2 + 10476103 n^4 + 23241063 n^6   \nn\\&&
 + 39815621 n^8+56151941 n^{10} + 68754245 n^{12} + 68754245 n^{14}),\nn\\
 \beta_9&=&\frac{(1 + n) (-1 + 2 n) (1 + 2 n)}{159659194286592000 n^{17}}(533203385 + 3285142432 n^2 + 10422722310 n^4\nn\\&& + 22636653380 n^6 + 
 37802268025 n^8 + 51431899764 n^{10} + 57961903280 n^{12} + 
 46369522624 n^{14}),\nn\\
 \beta_{10}&=&\frac{1+n}{28383856762060800000 n^{19}}(92382140214 - 78242942986 n^2 - 787141648501 n^4 \nn\\&&- 220291764901 n^6  + 
 5745141549269 n^8+ 20206109561669 n^{10} + 41774107453109 n^{12} \nn\\&&+ 
 64449433040309 n^{14} + 82551382505909 n^{16} + 82551382505909 n^{18}).\nn
\eea
The mutual information is
\be
I^{d=10}_{A_1,A_2}=\frac{1}{437580}G_{8,0}^{(d=10)}(z)-\frac{1}{2078505}G_{9,1}^{(d=10)}(z)+\frac{1}{5290740}G_{10,2}^{(d=10)}(z)+\cdots.
\ee
When $z<<1$, it is given by 
\be
I^{d=10}_{A_1,A_2}=\frac{1}{437580}z^8+\frac{2}{230945}z^9 + \frac{27}{1293292}z^{10}+\cdots.
\ee

\section{Conclusions and discussions}

In this paper, we computed the R\'enyi mutual information of free massless scalars in even dimensions $d=4,6,8,10$  to the first few orders. The computation relies on the OPE of spherical twist operator in terms of the primary fields in the replicated theory. The result can be written in terms of conformal partial waves. This allows us to study the behavior when the two spheres are near to each other and check the validity of the OPE method. Our results go beyond the ones in the literature. For example, in four dimension, the mutual information has been computed to order $z^2$ in \cite{Cardy:2013nua}, but now we are able to obtain the results up to order $z^4$. Notice that the coefficients of each $z^n$ are actually at the same order, so the higher order terms are not suppressed and become important when two spheres are near to each other. 

The leading contribution to the R\'enyi mutual information  in the small $z$ limit, i.e. the case that two spheres are relatively small compared with their distance, is always\cite{Faulkner:2013ana,Cardy:2013nua} 
\be
I^{(n),d}\sim z^{d-2} \sim \frac{1}{r^{2(d-2)}}. 
\ee
This is due to the fact that such contribution comes from the primary operators $\phi^2$ and $\phi_i\phi_j$, which is of the scaling dimension $d-2$. The next-to-leading contribution is of order $z^{d-1}$, partially from higher order expansion of the conformal partial wave of $G_{d-2,0}$ and partially  from the operators with dimension $d-1$ and spin $1$.  One remarkable point is that the spin 1 operator always gives negative contribution. The next-next-to-leading contribution is of order $z^d$. In higher dimension $d\geq 6$, such contribution comes partially from the stress tensor and its cousin. Moreover, for the leading contribution in the even dimension, there is  
\be
I^{d}_{LO}=\frac{\sqrt{\pi}\G(d-1)}{4^{d-1}\G(d-\frac{1}{2})}z^{d-2},
\ee
which is in perfect match with the result in \cite{Agon:2015ftl}. It was argued that such contribution could be universal. It would be interesting to investigate how much the mutual information could be universal determined\cite{work}.

The study in this work can be generalized to other cases. An obvious generalization is to odd dimensions. In \cite{Cardy:2013nua, Agon:2015twa} the mutual information for three dimensional free scalar has been studied. It would be interesting to apply our prescription to this case. Another interesting application is to the free fermion. We wish to come back to these cases in the future. 

\section*{Acknowledgments}
B.C. was in part supported by NSFC Grant No.~11275010, No.~11335012 and No.~11325522. J.L. is supported by the ERC Starting Grant 335146 ``HoloBHC".

\appendix
\section{Two-point functions in even dimensions}
We assume that the two fields $\phi(x),\phi(x')$ are at
\be
x=(r,\theta,\vec{y}),\ x'=(r',0,\vec{0}).
\ee
We define three quantities $\eta,\eta^{\pm}$
\be
\cosh\eta=\frac{r^2+r'^2+\vec{y}^2}{2rr'},\ \eta^{\pm}=\frac{\eta\pm i\theta}{2n}
\ee
to simplify notation. The the scalar two-point functions in the conical geometry in different even dimensions are the following. 
\begin{enumerate}
\item $d=4$
\be
<\phi(x)\phi(x')>_n=\frac{1}{4r r' n\sinh\eta}(\coth\eta^++\coth\eta^-).
\ee
\item $d=6$
\bea
\lefteqn{<\phi(x)\phi(x')>_n}\\
&=&\frac{1}{(4nrr'\sinh\eta)^2}(-2+\coth^2\eta^-+\coth^2\eta^++2n\coth\eta(\coth\eta^-+\coth\eta^+)).\nn
\eea
\item $d=8$
\bea
\lefteqn{<\phi(x)\phi(x')>_n}\nn\\
&=&\frac{1}{(4rr'n\sinh\eta)^3}(6n^2\coth^2\eta(\coth\eta^-+\coth\eta^+)\nn\\&&+3n\coth\eta(-2+\coth^2\eta^-+\coth^2\eta^+)+(\coth\eta^-+\coth\eta^+)\cdot\nn\\&&\cdot(-1-2n^2+\coth^2\eta^-+\coth^2\eta^+-\coth\eta^-\coth\eta^+)).
\eea
\item $d=10$
\bea
\lefteqn{<\phi(x)\phi(x')>_n}\nn\\
&=&\frac{1}{3\times(4rr'n\sinh\eta)^4}(2+16n^2-4(1+2n^2)\coth^2\eta^-+3\coth^4\eta^-\nn\\&&-4(1+2n^2)\coth^2\eta^++3\coth^4\eta^++60n^3\coth^3\eta(\coth\eta^-+\coth\eta^+)\nn\\&&-12n\coth\eta(\coth\eta^-+\coth\eta^+)
(1+3n^2-\coth^2\eta^--\coth^2\eta^++\nn\\&&+\coth\eta^-\coth\eta^+)+
30n^2\coth^2\eta(-2+\coth^2\eta^-+\coth^2\eta^+)).
\eea
\end{enumerate}

\section{Coefficients of one-point functions and normalizations}
In general dimension $d$, the one-point function of a primary operator in a conical geometry is fixed up to one coefficient $a_{\D,J}$. Assuming that the transverse and longitudinal  coordinates are respectively $x^a,y^i$, we list the convention of $a_{\D,J}$ for spin 0, 1 and 2 respectively
\be
<O(x)>_n=a_{\D,0}\frac{1}{|x|^{\D}},\ 
<O_a(x)>_n=a_{\D,1}\frac{\epsilon_{ab}n^b}{|x|^{\D}},\ <O_i(x)>_n=0,
\ee
\bea
&&<O_{ab}(x)>_n=-a_{\D,2}\frac{(d-1)\delta_{ab}-dn_an_b}{|x|^{\D}},\nn\\
&& <O_{ai}(x)>_n=0,\  <O_{ij}(x)>_n=a_{\D,2}\frac{\delta_{ij}}{|x|^{\D}}.
\eea
We have defined a normal vector 
\be
n^a=\frac{x^a}{|x|}.\nn
\ee
The normalization $N_{\D,J}$ are fixed by two-point function
\be
<O(x)O(x')>=\frac{N_{\D,0}}{(x-x')^{2\D}},
\ee
\be
<O_{\mu}(x)O_{\nu}(x')>=N_{\D,1}\frac{I_{\mu\nu}(x-x')}{(x-x')^{2\D}},
\ee
\be
<O_{\mu\nu}(x)O_{\rho\sigma}(x')>=N_{\D,2}\frac{I_{\mu\nu,\rho\sigma}(x-x')}{(x-x')^{2\D}},
\ee
where we defined 
\be
I_{\mu\nu}(x)=\delta_{\mu\nu}-2n_{\mu}n_{\nu},\ I_{\mu\nu,\rho\sigma}(x)=\frac{1}{2}(I_{\mu\rho}(x)I_{\nu\sigma}(x)+I_{\mu\sigma}(x)I_{\nu\rho}(x))-\frac{1}{d}\delta_{\mu\nu}\delta_{\rho\sigma}. \nn
\ee

In $d=6,8,10$, the primary operators, which have nonvanishing contribution to the R\'enyi mutual information and have scaling dimensions smaller than $d$, consist of 
\be
\phi_j^2, \hs{2ex}\phi_{j_1}\phi_{j_2},\hs{2ex}J_\m^{(j_1j_2)},\hs{2ex}T_{\m\n}^{(j)},\hs{3ex}T_{\m\n}^{(j_1j_2)}. \nn
\ee  
Their one-point function coefficients and normalization factors in different dimensions are listed in the following. In $d=6$, 
\bea
&&a_j=\frac{1+10n^2-11n^4}{720n^4},\hs{3ex} a_{ij}=\frac{(-2+2n^2+3\csc^2\frac{\theta_{ij}}{2n})\csc^2\frac{\theta_{ij}}{2n}}{48n^4},\nn\\
&&a_{J}=\frac{ (5 + 
   n^2 - (-1 + n^2) \cos\frac{\theta_{ij}}{n}) \cot\frac{\theta_{ij}}{2n}\csc^4\frac{\theta_{ij}}{2 n}}{24 n^5}.\nn\\
   &&a^{(j)}_T=\frac{10 + 21 n^2 - 31 n^6}{45360 n^6},\nn\\
   &&a^{(j_1j_2)}_T=-\frac{(-3(11+n^2)+2(-13+n^2)\cos\frac{\theta_{j_1j_2}}{n}+(-1+n^2)\cos\frac{2\theta_{j_1j_2}}{n})\csc^6\frac{\theta_{j_1j_2}}{2n}}{576n^6}\nn\\
   &&N_j=2,\ N_{j_1j_2}=1, N_J=8, N^{(j)}_T=\frac{160}{3},\ N^{(j_1j_2)}_T=\frac{80}{3}.\nn
\eea
In $d=8$, 
\bea
a&=&\frac{2+21n^2+168n^4-191n^6}{60480n^6},\nn\\ 
a_{ij}&=&\frac{(2-10n^2+8n^4+15(-1+n^2)\csc^2\frac{\theta_{ij}}{2n}+15\csc^4\frac{\theta_{ij}}{2n})\csc^2\frac{\theta_{ij}}{2n}}{960n^6}\nn\\
a_J&=&\frac{1}{1920 n^7}\left(3 (41 + 15 n^2 + 4 n^4) - 
   8 (-7 + 5 n^2 + 2 n^4) \cos\frac{\theta_{ij}}{n}\right.
     \nn\\&&\left.+(1 - 5 n^2 + 4 n^4) \cos\frac{2\theta_{ij}}{n}
     \right) \cot\frac{\theta_{ij}}{
  2 n} \csc^6\frac{\theta_{ij}}{2 n}\nn\\
  a^{(j)}_T&=&\frac{21 + 100 n^2 + 168 n^4 - 289 n^8}{2419200 n^8},\ \nn\\
  a_T^{(j_1j_2)}&=&\frac{1}{7680 n^8}\csc^2\frac{\theta_{j_1j_2}}{
  2 n} \left(-4 (1 - 5 n^2 + 4 n^4) + 
   6 (21 - 25 n^2 + 4 n^4) \csc^2\frac{\theta_{j_1j_2}}{
  2 n}\right. \nn\\&&
  \left. +30 (-14 + 5 n^2) \csc^4\frac{\theta_{j_1j_2}}{
  2 n} + 
   315 \csc^6\frac{\theta_{j_1j_2}}{
  2 n}\right)
\nn\\
N_j&=&2,\ N_{j_1j_2}=1,\ N_J=12,\ N^{(j)}_T=126,\ N_T^{(j_1j_2)}=63.\nn
\eea
In $d=10$, 
\bea
a&=&\frac{3+40n^2+294n^4+2160n^6-2497n^8}{3628800n^8},\nn\\
a_{ij}&=&\frac{1}{80640n^8}\csc^2\frac{\theta_{ij}}{2n}\left(4(-1+14n^2-49n^4+36n^6)+42(3-10n^2+7n^4)\csc^2\frac{\theta_{ij}}{2n}\right.\nn\\&&\left.+420(-1+n^2)\csc^4\frac{\theta_{ij}}{2n}+315\csc^6\frac{\theta_{ij}}{2n}\right),\nn\\
a_J&=&\frac{1}{322560 n^9 }\cot\frac{\theta_{ij}}{
  2 n} \csc^8\frac{\theta_{ij}}{2 n}
\left(5786 + 2660 n^2 + 1274 n^4 + 360 n^6 \right.\nn\\
& &- 3 (-1 + n) (1 + n) (1349 + 719 n^2 +
      180 n^4) \cos\frac{\theta_{ij}}{n} \nn\\&& \left.+ 
   6 (41 - 126 n^2 + 49 n^4 + 36 n^6) \cos\frac{2\theta_{ij}}{
     n} + (1 - 14 n^2 + 49 n^4 - 36 n^6) \cos\frac{
     3 \theta_{ij}}{n}\right),\nn\\
     a_T^j&=&\frac{30 + 231 n^2 + 770 n^4 + 1188 n^6 - 2219 n^{10}}{99792000 n^{10}},\nn\\
     a_T^{ij}&=&\frac{1}{201600 n^{10}}\csc^2\frac{\theta_{ij}}{2 n}(2 - 28 n^2 + 98 n^4 - 72 n^6 + 3 (-85 + 294 n^2 - 245 n^4 + 36 n^6)\csc^2\frac{\theta_{ij}}{2 n}+\nn\\&&735 (3 - 4 n^2 + n^4) \csc^4\frac{\theta_{ij}}{2 n} + 
  315 (-15 + 7 n^2)\csc^6\frac{\theta_{ij}}{2 n} + 
  2835 \csc^8\frac{\theta_{ij}}{2 n}),\nn\\
  N_j&=&2,\ N_{j_1j_2}=1,\ N_J=16,\ N^{(j)}_T=\frac{1152}{5}, \ N_T^{(j_1j_2)}=\frac{576}{5}.\nn
\eea

\section{Diagonal limit of conformal blocks}
The discussions on  conformal blocks in even dimensions can be found in \cite{Dolan:0011,Dolan:0309}. In this work, we consider the case $z=\bar{z}$, which is the diagonal limit of conformal blocks. 
The diagonal limit of conformal blocks in general dimension $d$ can be found in\cite{Matthijs:1305}. In our case 
\be
G_{\D,J}(z)=F_{\lambda,J}(Z), 
\ee
where $Z=\frac{z^2}{4(z-1)},\ \lambda=\frac{\D+J}{2}$. 
When $J$ is even, $J=2n$
\bea
F_{\lambda,2n}(Z)&=&(-4Z)^{\lambda-n}\frac{(\lambda-n+\frac{1}{2})_n}{(\lambda-n)_n(\lambda-\e-2n+\frac{1}{2})_n}\sum_{r=0}^nC_n^r\frac{(\frac{1}{2})_r(\e+n)_r(\l-\e-2n)_{n-r}}{(\l-n+\frac{1}{2})_r}\nn\\&&\times{}_3F_2(\l-n-\e,\l-n,\l-n+r,2\l-2n-\e,\l-n+r+\frac{1}{2};Z).
\eea
When $J=2n+1$ is odd, we have 
\bea
\lefteqn{F_{\l,2n+1}(Z)}\nn\\
&=&(\frac{1}{z}-\frac{1}{2})(-4Z)^{\l-n}\frac{(\lambda-n+\frac{1}{2})_n}{(\lambda-n)_n(\lambda-\e-2n-\frac{1}{2})_n}\sum_{r=0}^nC_n^r\frac{(\frac{1}{2})_r(\e+n+1)_r(\l-\e-2n-1)_{n-r}}{(\l-n+\frac{1}{2})_r}\nn\\&&\times{}_3F_2(\l-n-\e,\l-n,\l-n+r,2\l-2n-1-\e,\l-n+r+\frac{1}{2};Z).
\eea
Here we have defined $\e=\frac{d-2}{2}$. 

\section{Partition function}
According to \cite{Cardy1991,Cardy:2016lei,Dowker:2016twk},  the partition function for a real massless scalar including different components in dimension $d$ is 
\be
\tilde{Z}_d=\prod_{i=0}^{\infty}\frac{1}{(1-q^{d/2-1+i})^{D_d(i)}}, 
\ee
where 
\be
D_d(i)=\frac{(d+2i-2)(d+i-3)!}{i!(d-2)!}
\ee
is the dimension of the representation of $O(d)$ with angular momentum $i$.    In this work, the partition function $Z_d$ is to count different types of independent operators. The two partition function looks different, though they are actually related. We illustrate the difference in the case $d=4$. We expand $\tilde{Z}_d$ and $Z_d$ in the following
\bea
Z_4&=&1 + q + 2 q^2 + 3 q^3 + 5 q^4 + 7 q^5 + 11 q^6+\cdots,\nn\\
\tilde{Z}_4&=&1 + q + 5 q^2 + 14 q^3 + 40 q^4 + 101 q^5 + 266 q^6+\cdots.
\eea
At level $0$, there is only one independent operator $I$. This is the same for $Z_4, \tilde{Z}_4$ as the operator $I$ has one independent components. At level $1$, there is only one independent operator $\phi$. This is also the same for $Z_4, \tilde{Z}_4$ as the operator $\phi$ is just a scalar. At level $2$, there are two different types of operators $\phi^2,\partial_{\mu}\phi$ while there are totally $5$ different components.  Obviously, $\partial_{\mu}\phi$ is a descendant operator. So at level $2$, there are only one operator $\phi^2$ which is primary.  At level $3$, there are three different types of operators $\phi^3(1),\phi\partial_{\mu}\phi(4),\partial_{\mu}\partial_{\nu}\phi(9)$. The number in the bracket behind the operator counts the independent components. The total number is $14$ which is exactly what $\tilde{Z}_4$ counts.  Note among the three types of operators, $\phi\partial_{\mu}\phi,\partial_{\mu}\partial_{\nu}\phi$ are obviously descendant operators. So there are $1$ independent primary operator $\phi^3$ at level $3$.  At level $4$, there are five different types of operators $\phi^4(1),\phi^2\partial_{\mu}\phi(4),\phi\partial_{\mu}\partial_{\nu}\phi(9),\partial_{\mu}\phi\partial_{\nu}\phi(10),\partial_{\mu}\partial_{\nu}\partial_{\rho}\phi(16)$. The total number of independent components is $40$, exactly the same  number in $\tilde{Z}_4$. Note that $\phi^2\partial_{\mu}\phi, \partial_{\mu}\partial_{\nu}\partial_{\rho}\phi$ are descendant operators and  there are only 2 independent types of operators which are not descendant operators at this level, as suggested by $Z_4$. $\phi^4$ is independent and also primary. $\phi\partial_{\mu}\partial_{\nu}\phi$ and $\partial_{\mu}\phi\partial_{\nu}\phi$ should form one independent type of operator which is not descendant.  This is stress tensor $T_{\mu\nu}$. Note  the trace of the operators $\phi\partial_{\mu}\partial_{\nu}\phi$ and $\partial_{\mu}\phi\partial_{\nu}\phi$ are either descendant or zero on shell, so there is no additional primary operator. At level $5$, there are seven different types of operators $\phi^5(1),\phi^3\partial_{\mu}\phi(4),\phi^2\partial_{\mu}\partial_{\nu}\phi(9),\phi\partial_{\mu}\phi\partial_{\nu}\phi(10),\phi\partial_{\mu}\partial_{\nu}\partial_{\rho}\phi(16),\partial_{\mu}\phi\partial_{\nu}\partial_{\rho}\phi(36),\partial_{\mu}\partial_{\nu}\partial_{\rho}\partial_{\sigma}\phi(25)$, among which $\phi^3\partial_{\mu}\phi,\phi\partial_{\mu}\partial_{\nu}\partial_{\rho}\phi,\partial_{\mu}\phi\partial_{\nu}\partial_{\rho}\phi,\partial_{\mu}\partial_{\nu}\partial_{\rho}\partial_{\sigma}\phi$ are descendent operators. $Z_4$ at level 5 tells us that there are only two independent types of operators which are not descendants. The first one is $\phi^5$ which is also primary. The other one is a combination of $\phi^2\partial_{\mu}\partial_{\nu}\phi, \phi\partial_{\mu}\phi\partial_{\nu}\phi$. To use these operators to construct the primary operators, one should decompose them into irreducible representation of $SO(4)$, say the symmetric traceless one.  As the trace of these operators are either descendants or zero, the only possible primary operator is a spin $2$ operator $\phi T_{\mu\nu}$.  

From the above discussion we see that the definition of $Z_d$ is quite useful to study the primary operators while $\tilde{Z}_d$ is not convenient. In general $m(k)=p(k)-p(k-1)$ gives the number of independent types of operators\footnote{$k=0,1$ should be considered separately.} which are not descendants. Based on these non-descendant operators, we can construct the primary operators in the irreducible representation of $SO(4)$. Technically  one should check whether one can construct more primary operators by taking trace of the candidate operators\footnote{There may be primary operators with mixed symmetry, but we do not find them at least to the order we considered in this work.}. For $k\le 6$, we find that all trace terms are either descendants or zero in $d=4$. So $m(k)$ is just the number of independent primary operators for $k\le 6$.  We summarize the discussion in the following table. To construct the primary operator at level $k$, one first selects all possible independent operators which are not descendants. There should be $m(k)$ terms. These operators are candidate operators. Then one should organize these operators into the irreducible representation of $SO(d)$ to find the primary operators. 
\begin{table}
 \centering
\begin{tabular}{|c|c|c|c|}
\hline
$\Delta$&operators&number&components\\
\hline 0&$I$&$1$&$1$\\
\hline 1&$\phi$&$1$&$1$\\
\hline 2&$\phi^2,\partial_{\mu}\phi$&$2$&$5$\\
\hline 3&$\phi^3,\phi\partial_{\mu}\phi,\partial_{\mu}\partial_{\nu}\phi$&$3$&$14$\\
\hline 4&$\phi^4,\phi^2\partial_{\mu}\phi,\partial_{\mu}\phi\partial_{\nu}\phi,\phi\partial_{\mu}\partial_{\nu}\phi,\partial_{\mu}\partial_{\nu}\partial_{\rho}\phi$&$5$&$40$\\
\hline 5&$\phi^5,\phi^3\partial_{\mu}\phi,\phi^2\partial_{\mu}\partial_{\nu}\phi,\phi\partial_{\mu}\phi\partial_{\nu}\phi,\phi\partial_{\mu}\partial_{\nu}\partial_{\rho}\phi,\partial_{\mu}\phi\partial_{\nu}\partial_{\rho}\phi,\partial_{\mu}\partial_{\nu}\partial_{\rho}\partial_{\sigma}\phi$&$7$&$101$\\
\hline
\end{tabular}
\caption{Check the consistency of $Z_d$ and $\tilde{Z}_d$ in $d=4$}
\end{table}

For the replicated theory, the partition functions can be defined by \be
Z_{n,d}=(Z_d)^n, \hs{3ex}\tilde{Z}_{n,d}=(\tilde{Z}_d)^n.
\ee
We  check that the two partition functions  are consistent up to level $d$. The primary operators can be constructed systematically.

\end{document}